\documentstyle{article}

\begin{document}
\title{Quantum electrodynamics in the squeezed vacuum state:
Electron mass shift}
\author{
  V. Putz$^1$ and K. Svozil$^2$,\\
  {\small$^1$ Max-Planck-Institute for Mathematics in the Sciences}\\
  {\small Inselstra\ss e 22-26, D-04103 Leipzig, Germany}\\
  {\small Phone: +49 - 341 - 9959 - 676}\\
  {\small Fax: +49 - 341 - 9959 - 658} \\
  {\small vputz@mis.mpg.de}\\ \\
  {\small$^2$ Institut f\"ur Theoretische Physik,
  Technische Universit\"at Wien   }     \\
  {\small Wiedner Hauptstra\ss e 8-10/136,    
  A-1040 Vienna, Austria   }            \\
  {\small svozil@tuwien.ac.at}}
\maketitle
\footnotetext[1]{work supported in part by ``Fonds zur F\"orderung der 
Wissenschaftlichen Forschung'' (FWF) under contract P15015-TPH}  
\newpage
\begin{abstract}
Due to the nonvanishing average photon population of the squeezed vacuum
state, finite corrections to the scattering matrix are obtained.  The
lowest order contribution to the electron mass shift for a one mode
squeezed vacuum state is given by $\delta m(\Omega ,s)/m=\alpha (2/\pi
)(\Omega /m)^2\sinh^2(s)$, where $\Omega$ and $s$ stand for the mode
frequency and the squeeze parameter and $\alpha$ for the fine structure
constant, respectively.
\end{abstract}
\vspace{1cm}
PACS-1996: \\
11.10.Wx Finite-temperature field theory\\
12.20.-m  Quantum electrodynamics\\ \\

The squeezed vacuum is a fascinating nonclassical state of the
quantized electromagnetic field \cite{loudon}.
Just as for the finite temperature case,
the squeezed vacuum is populated by
photons.
Therefore, the scattering matrix, and in particular renormalization,
has to be re-evaluated with these finite ground state photons in mind.

The dependence of the scattering matrix on the vacuum state
of the theory and on exterior parameters has been studied previously for the
 thermal
equilibrium \cite{thermal}, in cavity--quantum
electrodynamics \cite{cavity-qed},
on fractal space--time support \cite{zei-svo}
and, to some extent, in the presence of strong electromagnetic
fields \cite{greiner,strongfield-qed}.
Here, quantum electrodynamics
is investigated
in the presence of squeezed vacuum fluctuations \cite{milburn}; i.e.,
 fluctuations with reduced noise in amplitude or phase.

At first we shall calculate the scattering matrix by Taylor-expansion up to
second order of $e^{2}$. Let $\vert{\rm i}\rangle =
a_{e}^{(r)\dagger}(\vec q)\vert{\rm sv}\rangle$ be the initial state, $r$ the incoming
electron's spin, $\vec q$ its momentum and $\vert{\rm sv}\rangle$ the squeezed
vacuum state. $\vert{\rm sv}\rangle$  is a pure photonic state and behaves like an ordinary
Fock-vacuum regarding the electron creation and annihilation operators.
The final state is $\langle{\rm f}\vert =
\langle{\rm sv}\vert a_{e}^{(r')}(\vec q')$. It is important to remark that
the initial squeezed vacuum state will be assumed to be the same as the final
one. Hence, in this approximation, $\vert{\rm sv}\rangle$ is time independent.

The scattering matrix is given by
 \begin{eqnarray*}
 \langle{\rm f}\vert\rm S\vert{\rm i}\rangle&=&\langle{\rm f}\vert\rm {T}
 e^{i\int{d^4x\mathcal{L_W}}}\vert{\rm i}\rangle,
 \end{eqnarray*}
where
$\mathcal{L_W}$ stands for the interaction-term of the Lagrange-density, which, in
QED is is given by ${\mathcal L_W} = -e:\!\overline \psi\rlap{/}A\psi \!:$.
The expansion of $S$ with respect to $e$ is
 \begin{eqnarray*}
 &&S \approx 1 - ie\int{d^4x:\!\overline\psi(x) \rlap{/}A(x)\psi(x)\!:} +\\
&& \qquad
 {(-ie)^2\over(2!)} \int\!\!\!\int{d^4xd^4y\rm{T}[:\!\overline\psi(x) \rlap{/}A(x)\psi(x)\!:
 :\!\overline\psi(y)(\rlap{/}A(y)\psi(y)\!:]}.
 \end{eqnarray*}
 We shall discuss the first three terms in the series expansion in $e$ next.
 We find
 \begin{eqnarray*}
 &&{\mathcal O}(e^0): \langle{\rm f}\vert 1\vert{\rm i}\rangle=\delta^3(\vec q -
 \vec q')\delta_{rr'}\\
 \\
 &&{\mathcal O}(e^1): \langle{\rm f}\vert(-ie\int{d^4x:\!\overline\psi(x)\rlap{/}A(x)\psi(x)\!:)
 \vert{\rm i}\rangle} = 0.
 \end{eqnarray*}
$A_{\mu}(x)$ contains a term with exactly one annihilation
operator and a term with exactly one creation operator, so that
$\langle{\rm f}\vert a^{(\dagger)}\vert{\rm i}\rangle = 0$.
The well known relations $\langle{\rm sv}\vert a\vert{\rm sv}\rangle = 0$
and $\langle{\rm sv}\vert a^\dagger\vert{\rm sv}\rangle = 0$ hold.

${\mathcal O}(e^2):$ The electron- and photon-operators do not act on each
other. Hence they commute and $\vert{\rm sv}\rangle$ is a normal Fock-vacuum
for the electron operators. Therefore it is possible to completely separate the electron
and photon terms
 \begin{eqnarray*}
 &&\langle {\rm f} \vert {(-ie)^2\over(2!)} \int \!\!\! \int d^4xd^4y {\rm T}
 [:\!\overline\psi(x)\rlap{/}A(x)\psi(x)\!::\!\overline\psi(y)\rlap{/}A(y)\psi(y)\!:]
 \vert {\rm i} \rangle =\\
&&={-e^2\over 2} \int \!\!\! \int d^4xd^4y{\rm T}\langle{\rm sv}
 \vert A_{\mu}(x)A_{\nu}(y)\vert {\rm sv}\rangle\times\\&&\quad\times
 {\rm T}\langle\rm{0}
 \vert a_{e}^{(r')}(\vec q'):\!\overline \psi(x)\gamma^{\mu} \psi(x)\!::\!\overline
 \psi(y)\gamma^{\nu} \psi(y)\!:a_{e}^{(r)\dagger}(\vec q)\vert{\rm 0}\rangle.
 \end{eqnarray*}
The electron term is given by
 \begin{eqnarray*}
 &&\underbrace{\delta(\vec q - \vec q')\delta_{rr'}\langle{\rm 0}\vert:\!
 \overline \psi(x)\gamma^{\mu} \psi(x)\!::\!\overline \psi(y)\gamma^{\nu}
 \psi(y)\!:\vert{\rm 0}\rangle}_{disconnected} + \\
&&+
 \underbrace{{e^{iq'x}\over(2\pi)^{3/2}\sqrt{2\overline q'^{0}}}\overline
 u^{(r')}\gamma^{\mu}iS_{c}(x-y)\gamma^{\nu}{e^{-iqy}\over(2\pi)^{3/2}
 \sqrt{2\overline q^{0}}} u^{(r)} +(x\leftrightarrow y,\mu
 \leftrightarrow \nu)}_{connected}.
 \end{eqnarray*}
As usual, the disconnected-term is regarded as nonphysical.

The calculation
demonstrated that effectively it would have been possible to build up the whole 2nd-order term
 by just replacing the usual photon propagator in the Feynman-rules by
$$iD_{\mu\nu}(x-y)=\langle{\rm sv}\vert{\rm T}[A_{\mu}(x)A_{\nu}(y)]
\vert{\rm sv}\rangle.$$
This expression can be evaluated as follows:
\begin{eqnarray*}
&& \langle {\rm sv}\vert T[A_\mu (x)A_\nu (y)]\vert {\rm sv}\rangle= 
{1\over (2\pi)^3} \int \!\!\!\int {d^3kd^3k'\over2(E_k E_{k'})^{1/2}}\\&&\quad
 \langle{\rm sv}\vert \theta(x_0 - y_0)[\epsilon^{(\rho)}_{\mu}(\vec k)
 a^{-}_{\rho}(\vec k)\epsilon^{(\lambda)}_{\nu}(\vec k')a^{\dagger}_{\lambda}
 (\vec k')e^{-i(kx-k'y)} \\
 &&\quad +\epsilon^{(\rho)}_{\mu}(\vec k)a^{\dagger}_{\rho}(\vec k)
 \epsilon^{(\lambda)}_{\nu}(\vec k')a^{-}_{\lambda}
 (\vec k')e^{i(kx-k'y)}] + (x \leftrightarrow y)\vert {\rm sv}\rangle ,
\end{eqnarray*}
 where
  \begin{eqnarray*}
&& \langle {\rm sv}\vert a^{\dagger }_{\rho }(\vec k)a^{-}_{\lambda }(\vec k')\vert {\rm sv }\rangle
=
-g_{\rho \lambda }\delta^3(\vec k-\vec k')n(k),\\
&&[ a^{-}_{\rho }({\vec k}) , a ^ {\dagger } _ {\lambda }({\vec k}')]
  =
-g_{\rho \lambda }\delta^3(\vec k-\vec k'),\\
 &&g_{\rho \lambda }\epsilon^{(\rho )}_{\mu}(\vec k)\epsilon^{(\lambda )}_{\nu }(\vec k)  =  g_{\mu \nu}.
\end{eqnarray*}
Then,
\begin{eqnarray*}
&& \langle {\rm sv}\vert T[A_\mu (x)A_\nu (y)]\vert {\rm sv}\rangle= \\
 &&= -g_{\mu \nu}\int {d^3k\over (2\pi )^3 2E_k}
 [\theta (x_0-y_0)e^{-ik(x-y)}+
 \theta (y_0-x_0)e^{ik(x-y)}]  \\&&\quad
 -g_{\mu \nu}\int {d^3k\over (2\pi )^3 2E_k}
  n(k)[\theta (x_0-y_0)e^{-ik(x-y)} +
 \theta (x_0-y_0)e^{ik(x-y)} \\&&\qquad\qquad+ \theta (y_0-x_0)e^{-ik(y-x)} +
 \theta (y_0-x_0)e^{ik(y-x)}]          .
\end{eqnarray*}
Hence we obtain
\begin{eqnarray}
 &&iD_{\mu \nu}(x-y)=\langle {\rm sv}\vert T[A_\mu (x)A_\nu (y)]\vert
 {\rm sv}\rangle \nonumber \\
 &&=-g_{\mu \nu}\left\{ \int {d^3k\over (2\pi )^3}{1\over 2E_k}[\theta
 (x_0-y_0)e^{-ik (x-y)}+\theta (y_0-x_0)e^{ik (x-y)}]+
 \right.
 \nonumber \\
 &&\qquad \qquad +  \left.\int
 {d^3k\over (2\pi )^3}{1\over 2E_k}
 n(k)[e^{ik (x-y)}+e^{-ik (x-y)}]\right\}
.
 \end{eqnarray}

Notice, as remarked above, that by defining the photon propagator,
the squeezed vacuum state had to be assumed ``quasi-stationary,''
otherwise the final state of the vacuum cannot be identified with the
initial state. (This assumption can be justified only within the appropriate
spatial and temporal ranges.)
The propagator can be rewritten using contour--integral techniques
\begin{eqnarray}
iD_{\mu \nu}(x-y)&=&i\int {d^4k\over (2\pi )^4}e^{-ik(x-y)}D_{\mu \nu }
 (k)\nonumber \\
iD_{\mu \nu }(k)&=&-ig_{\mu \nu }\left[{1\over k^2+i\epsilon }-2\pi i
\delta (k^2)n(k)\right].                     \label{prop}
\end{eqnarray}
For the one mode squeezed state, $n(k;\Omega ,s)=\Omega \sinh^2(s)
\delta (
E_k -\Omega )$, where $E_k$ is the photon energy parameter and $\Omega$
 and $s$ stand for the frequency of
the squeezed mode and the squeezing parameter, respectively.
The  electron propagator $S(p)=1/(\rlap{/}p -m+i\epsilon )$, as well
as the bare vertex $\gamma_\mu$ remain unchanged.
 Notice however that a preferred frame of
 reference has been introduced due to the noncovariant choice of the
 density $n(k;\Omega ,s)$, i.e., the one at rest with respect to
 the squeezed vacuum.

In what follows, the lowest order correction to the radiative mass of
the electron will be calculated. This can be done by evaluating
the second order contribution to the self energy
of the electron
\begin{equation}
 -i\Sigma (p;\Omega ,s)=\int{d^4k\over (2\pi )^4}[iD_{\mu
\nu}(k;\Omega ,s)](-ie)\gamma^\mu {i\over \rlap{/}p-\rlap{/}k-m}(-ie)\gamma^\nu
.
\end{equation}
The physical mass is interpreted  as the pole of the
renormalized electron propagator. For $\delta m(\Omega ,s)\ll m$,
\begin{eqnarray}
m(\Omega ,s)&\approx &m-\delta m+\Sigma (p;\Omega
 ,s)\vert_{\rlap{/}p=m}\nonumber \\
 & &\qquad =m-\delta m+\Sigma
(p;s=0)\vert_{\rlap{/}p=m}+\delta \Sigma
(p;\Omega,s)\vert_{\rlap{/}p=m}\nonumber \\
 & &\qquad =m+\delta m(\Omega ,s),
\end{eqnarray}
where $m$ stands for the renormalized nonsqueezed mass of the electron.

The correction term $\delta m(\Omega ,s) =\delta
\Sigma(p;\Omega ,s)\vert_{\rlap{/}p=m}$
due to squeezing adds up coherently to the renormalization
contributions of $m$. Its explicit form is given by

\begin{eqnarray*}
 \delta m(\Omega ,s)&=&-{e^2\over (2\pi )^3}\int d^4k \delta
 (k^2)n(k;\Omega ,s)\gamma_\mu
 {\rlap{/}p-\rlap{/}k+m\over  (p-k)^2-m^2+i\epsilon }\gamma^\mu
 \mid_{\rlap{/}p=m}  \\
 &=&\int{d^4k}\delta (k^2)n(k)\gamma_{\mu}{\rlap{/}p-\rlap{/}k+m\over (p-k)^2
 -m^2+i\epsilon}\gamma^{\mu}\vert_{\rlap{/}p=m} \\
 &=& -2\int{d^4k}\delta(k^2)n(k){\rlap{/}k+m\over 2pk-i\epsilon}
 \vert_{\rlap{/}p=m}\\
  &=&-\int{d^3\vec k dk^0}[{\delta(k^0-\vert\vec k\vert)\over\vert 2k^0\vert}
 + {\delta(k^0+\vert\vec k\vert)\over\vert 2k^0\vert}]n(k){k^0\gamma_0-
 \vec k \vec\gamma + m\over k^0 p_0-\vec k\vec p-i\epsilon}\vert_{\rlap{/}p=m} .
\end{eqnarray*}
As the epsilon is not needed, it will be dropped,
\begin{eqnarray}
 &&= -\int{d^3k} n(\vert\vec k\vert)[{\vert\vec k\vert\gamma_0-\vec k\vec\gamma
 +2m \over 2 \vert\vec k\vert(\vert\vec k\vert - \vec k\vec p)}+
 \underbrace{{-\vert\vec k\vert\gamma_0-\vec k\vec \gamma + 2m  \over 2 \vert\vec k\vert(-\vert\vec k\vert p_0-\vec k\vec p)}}
 _{\vec k \rightarrow -k}]\vert_{\rlap{/}p=m} \nonumber \\
 &&= -\int {d^3k} n(\vert\vec k\vert)[{\vert\vec k\vert\gamma_0-\vec k\vec\gamma
 \over \vert\vec k\vert(\vert\vec k\vert p_0-\vec k\vec p)}\vert_{\rlap{/}p=m}  \nonumber  \\
 &&= -\int{d^3k} n(\vert\vec k\vert){k_{\mu}\gamma^{\mu}\over\vert\vec k\vert(pk)}
 \vert_{\rlap{/}p=m,\ e.o.m.:\ k^2=0} \nonumber \\
 &&={\alpha \over 2\pi^2}{I_\mu (p)p^\mu \over m}\vert_{p^2=m^2},
\end{eqnarray}
 where $\delta (k^2) = \delta(k^0-\vert \vec {\rm k}\vert)/2k^0 +
 \delta(k^0+\vert \vec {\rm k}\vert)/2k^0$ and
 Gordon's identity, which reduces to $\gamma_\mu =p_\mu /m$
 (remind $p_\mu \gamma^\mu = m,\ p^2 = m^2$), have been used,
 $\alpha =e^2/4\pi$ stands for the fine structure constant and
\begin{equation}
 I_\mu (p)=\int {d^3\vec {\rm k}}{k_\mu
 \over  \vert \vec  {\rm k}\vert (pk)}n(\vert \vec {\rm k}\vert ;\Omega
 ,s) \vert_{e.o.m.:\ k^2=0}.
 \end{equation}
In the rest frame of the squeezed vacuum
this expression can be  evaluated, yielding
\begin{equation}
{\delta m(\Omega ,s)/ m}= \alpha (2 / \pi)
(\Omega / m)^2\sinh^2(s).
\end{equation}
For optical frequencies, $\delta m(s)/m\approx 10^{-13}\sinh^2(s)$.

One has to bear in mind that the above calculation did {\em not} take
into explicit account the spatial and temporal characteristics of the squeezed
vacuum states. Therefore, a more careful
calculation would have to take into account the {\em nonstationary}
property of the squeezed vacuum.

However, even the above rather simple model calculations suggest
that physical parameters such as  electron mass, charge and magnetic moment
dependent on external conditions.
The squeezed vacuum is arguably the simplest theoretically treatable yet
experimentally realizable state.
Here, we have just
evaluated the electron-mass-shift.
Measuring the renormalization effects
on the electron mass due to the squeezed vacuum
is certainly a challenging yet difficult task beyond the scope of this presentation.
Calculations of charge-shift
and of corrections to the magnetic moment will be presented in a forthcoming paper.


\begin{thebibliography}{99}
\bibitem{fn1}
 other squeezing parameters associated with more general canonical
 transformations [isomorphic to elements of the real symplectic
 group $Sp(2N,R)$ for $N$--mode squeezing]
 have been set zero, but a generalization is
 straightforward.
\end{thebibliography}
\end{document}